\input amssym.tex
\input epsf
\epsfclipon


\magnification=\magstephalf
\hsize=14.0 true cm
\vsize=19 true cm
\hoffset=1.0 true cm
\voffset=2.0 true cm

\abovedisplayskip=12pt plus 3pt minus 3pt
\belowdisplayskip=12pt plus 3pt minus 3pt
\parindent=1.0em


\font\sixrm=cmr6
\font\eightrm=cmr8
\font\ninerm=cmr9

\font\sixi=cmmi6
\font\eighti=cmmi8
\font\ninei=cmmi9

\font\sixsy=cmsy6
\font\eightsy=cmsy8
\font\ninesy=cmsy9

\font\sixbf=cmbx6
\font\eightbf=cmbx8
\font\ninebf=cmbx9

\font\eightit=cmti8
\font\nineit=cmti9

\font\eightsl=cmsl8
\font\ninesl=cmsl9

\font\sixss=cmss8 at 8 true pt
\font\sevenss=cmss9 at 9 true pt
\font\eightss=cmss8
\font\niness=cmss9
\font\tenss=cmss10

 at 12 true pt
 at 12 true pt
\font\bigrm=cmr10 at 12 true pt
 at 12 true pt
 at 12 true pt

\font\Bigi=cmmi12 at 16 true pt
 at 16 true pt
 at 16 true pt
 at 16 true pt
\font\Bigrm=cmr12 at 16 true pt
 at 16 true pt
 at 16 true pt

\catcode`@=11
\newfam\ssfam

\def\tenpoint{\def\rm{\fam0\tenrm}%
    \textfont0=\tenrm \scriptfont0=\sevenrm \scriptscriptfont0=\fiverm
    \textfont1=\teni  \scriptfont1=\seveni  \scriptscriptfont1=\fivei
    \textfont2=\tensy \scriptfont2=\sevensy \scriptscriptfont2=\fivesy
    \textfont3=\tenex \scriptfont3=\tenex   \scriptscriptfont3=\tenex
    \textfont\itfam=\tenit                  \def\it{\fam\itfam\tenit}%
    \textfont\slfam=\tensl                  \def\sl{\fam\slfam\tensl}%
    \textfont\bffam=\tenbf \scriptfont\bffam=\sevenbf
    \scriptscriptfont\bffam=\fivebf
                                            \def\bf{\fam\bffam\tenbf}%
    \textfont\ssfam=\tenss \scriptfont\ssfam=\sevenss
    \scriptscriptfont\ssfam=\sevenss
                                            \def\ss{\fam\ssfam\tenss}%
    \normalbaselineskip=13pt
    \setbox\strutbox=\hbox{\vrule height8.5pt depth3.5pt width0pt}%
    \let\big=\tenbig
    \normalbaselines\rm}

\def\ninepoint{\def\rm{\fam0\ninerm}%
    \textfont0=\ninerm      \scriptfont0=\sixrm
                            \scriptscriptfont0=\fiverm
    \textfont1=\ninei       \scriptfont1=\sixi
                            \scriptscriptfont1=\fivei
    \textfont2=\ninesy      \scriptfont2=\sixsy
                            \scriptscriptfont2=\fivesy
    \textfont3=\tenex       \scriptfont3=\tenex
                            \scriptscriptfont3=\tenex
    \textfont\itfam=\nineit \def\it{\fam\itfam\nineit}%
    \textfont\slfam=\ninesl \def\sl{\fam\slfam\ninesl}%
    \textfont\bffam=\ninebf \scriptfont\bffam=\sixbf
                            \scriptscriptfont\bffam=\fivebf
                            \def\bf{\fam\bffam\ninebf}%
    \textfont\ssfam=\niness \scriptfont\ssfam=\sixss
                            \scriptscriptfont\ssfam=\sixss
                            \def\ss{\fam\ssfam\niness}%
    \normalbaselineskip=12pt
    \setbox\strutbox=\hbox{\vrule height8.0pt depth3.0pt width0pt}%
    \let\big=\ninebig
    \normalbaselines\rm}

\def\eightpoint{\def\rm{\fam0\eightrm}%
    \textfont0=\eightrm      \scriptfont0=\sixrm
                             \scriptscriptfont0=\fiverm
    \textfont1=\eighti       \scriptfont1=\sixi
                             \scriptscriptfont1=\fivei
    \textfont2=\eightsy      \scriptfont2=\sixsy
                             \scriptscriptfont2=\fivesy
    \textfont3=\tenex        \scriptfont3=\tenex
                             \scriptscriptfont3=\tenex
    \textfont\itfam=\eightit \def\it{\fam\itfam\eightit}%
    \textfont\slfam=\eightsl \def\sl{\fam\slfam\eightsl}%
    \textfont\bffam=\eightbf \scriptfont\bffam=\sixbf
                             \scriptscriptfont\bffam=\fivebf
                             \def\bf{\fam\bffam\eightbf}%
    \textfont\ssfam=\eightss \scriptfont\ssfam=\sixss
                             \scriptscriptfont\ssfam=\sixss
                             \def\ss{\fam\ssfam\eightss}%
    \normalbaselineskip=10pt
    \setbox\strutbox=\hbox{\vrule height7.0pt depth2.0pt width0pt}%
    \let\big=\eightbig
    \normalbaselines\rm}

\def\tenbig#1{{\hbox{$\left#1\vbox to8.5pt{}\right.\n@space$}}}
\def\ninebig#1{{\hbox{$\textfont0=\tenrm\textfont2=\tensy
                       \left#1\vbox to7.25pt{}\right.\n@space$}}}
\def\eightbig#1{{\hbox{$\textfont0=\ninerm\textfont2=\ninesy
                       \left#1\vbox to6.5pt{}\right.\n@space$}}}

\font\sectionfont=cmbx10
\font\subsectionfont=cmti10

\def\figurecaptionfont{\ninepoint}
\def\tablecaptionfont{\ninepoint}
\def\footnotefont{\eightpoint}


\newcount\equationno
\newcount\bibitemno
\newcount\figureno
\newcount\tableno

\equationno=0
\bibitemno=0
\figureno=0
\tableno=0


\footline={\ifnum\pageno=0{\hfil}\else
{\hss\rm\the\pageno\hss}\fi}


\def\section #1. #2 \par
{\vskip0pt plus .10\vsize\penalty-100 \vskip0pt plus-.10\vsize
\vskip 1.6 true cm plus 0.2 true cm minus 0.2 true cm
\global\def\equationlabel{#1}
\global\equationno=0
\leftline{\sectionfont #1. #2}\par
\immediate\write\terminal{Section #1. #2}
\vskip 0.7 true cm plus 0.1 true cm minus 0.1 true cm
\noindent}


\def\subsection #1 \par
{\vskip0pt plus 0.8 true cm\penalty-50 \vskip0pt plus-0.8 true cm
\vskip2.5ex plus 0.1ex minus 0.1ex
\leftline{\subsectionfont #1}\par
\immediate\write\terminal{Subsection #1}
\vskip1.0ex plus 0.1ex minus 0.1ex
\noindent}


\def\appendix #1. #2 \par
{\vskip0pt plus .20\vsize\penalty-100 \vskip0pt plus-.20\vsize
\vskip 1.6 true cm plus 0.2 true cm minus 0.2 true cm
\global\def\equationlabel{\hbox{\rm#1}}
\global\equationno=0
\leftline{\sectionfont Appendix #1. #2}\par
\immediate\write\terminal{Appendix #1. #2}
\vskip 0.7 true cm plus 0.1 true cm minus 0.1 true cm
\noindent}



\def\equation#1{$$\displaylines{\qquad #1}$$}
\def\enum{\global\advance\equationno by 1
\hfill\llap{{\rm(\equationlabel.\the\equationno)}}}
\def\noenum{\hfill}
\def\next#1{\cr\noalign{\vskip#1}\qquad}


\def\ifundefined#1{\expandafter\ifx\csname#1\endcsname\relax}

\def\ref#1{\ifundefined{#1}?\immediate\write\terminal{unknown reference
on page \the\pageno}\else\csname#1\endcsname\fi}

\newwrite\terminal
\newwrite\bibitemlist

\def\bibitem#1#2\par{\global\advance\bibitemno by 1
\immediate\write\bibitemlist{\string\def
\expandafter\string\csname#1\endcsname
{\the\bibitemno}}
\item{[\the\bibitemno]}#2\par}

\def\beginbibliography{
\vskip0pt plus .15\vsize\penalty-100 \vskip0pt plus-.15\vsize
\vskip 1.2 true cm plus 0.2 true cm minus 0.2 true cm
\leftline{\sectionfont References}\par
\immediate\write\terminal{References}
\immediate\openout\bibitemlist=biblist
\frenchspacing\parindent=1.8em
\vskip 0.5 true cm plus 0.1 true cm minus 0.1 true cm}

\def\endbibliography{
\immediate\closeout\bibitemlist
\nonfrenchspacing\parindent=1.0em}


\def\figurecaption#1{\global\advance\figureno by 1
\narrower\figurecaptionfont
Fig.~\the\figureno. #1}

\def\tablecaption#1{\global\advance\tableno by 1
\vbox to 0.5 true cm { }
\centerline{\tablecaptionfont%
Table~\the\tableno. #1}
\vskip-0.4 true cm}

\def\thintablerule{\hrule height0.4pt}

\tenpoint

\immediate\openin\bibitemlist=biblist
\ifeof\bibitemlist\immediate\closein\bibitemlist
\else\immediate\closein\bibitemlist
\input biblist \fi


\def\thismonth{\ifcase\month\or
January\or February\or March\or April\or May\or June\or
July\or August\or September\or October\or November\or December\fi}



\def\rmd{{\rm d}}

\def\rme{{\rm e}}



\def\proof{\noindent{\sl Proof:}\kern0.6em}

\def\frac#1#2{\hbox{$#1\over#2$}}
\def\dual{\mathstrut^*\kern-0.1em}

\def\lvec#1{\setbox0=\hbox{$#1$}
    \setbox1=\hbox{$\scriptstyle\leftarrow$}
    #1\kern-\wd0\smash{
    \raise\ht0\hbox{$\raise1pt\hbox{$\scriptstyle\leftarrow$}$}}
    \kern-\wd1\kern\wd0}
\def\rvec#1{\setbox0=\hbox{$#1$}
    \setbox1=\hbox{$\scriptstyle\rightarrow$}
    #1\kern-\wd0\smash{
    \raise\ht0\hbox{$\raise1pt\hbox{$\scriptstyle\rightarrow$}$}}
    \kern-\wd1\kern\wd0}


\def\nab#1{{\nabla_{#1}}}
\def\nabstar#1{{\nabla\kern0.5pt\smash{\raise 4.5pt\hbox{$\ast$}}
               \kern-5.5pt_{#1}}}

\def\drvstar#1{{\partial\kern0.5pt\smash{\raise 4.5pt\hbox{$\ast$}}
               \kern-6.0pt_{#1}}}

\def\ldrvstar#1{{\lvec{\,\partial}\kern-0.5pt\smash{\raise 4.5pt\hbox{$\ast$}}
               \kern-5.0pt_{#1}}}


\def\MeV{{\rm MeV}}

\def\fm{{\rm fm}}



\def\psibar{\overline{\psi}}


\def\dirac#1{\gamma_{#1}}
\def\diracstar#1#2{
    \setbox0=\hbox{$\gamma$}\setbox1=\hbox{$\gamma_{#1}$}
    \gamma_{#1}\kern-\wd1\kern\wd0
    \smash{\raise4.5pt\hbox{$\scriptstyle#2$}}}


\def\SUthree{{\rm SU(3)}}

\def\tr{{\rm tr}}
\def\Tr{{\rm Tr}}
\def\Ad{{\rm Ad}\kern0.1em}


\def\ubar{\bar{u}}

\def\sbar{\bar{s}}


\def\Ohat{{\hat{\cal O}}}


\def\Dw{D_{\rm w}}
\def\abar{\bar{a}}
\def\Dzp{D^{+}}
\def\Dzm{D^{-}}
\def\Dzpm{D^{\pm}}
\def\Dtildepm{\widetilde{D}^{\pm}}

\def\Dm{D_m}
\def\Dmdag{\setbox0=\hbox{$\displaystyle D$}%
           \setbox1=\hbox{$\displaystyle D_m$}%
           D_m\kern-\wd1\kern\wd0%
           \smash{\raise4.5pt\hbox{\kern0pt$\scriptstyle\dagger$}}\kern4pt}


\def\Pp{P_{+}}
\def\Pm{P_{-}}


\def\Pr{{\Bbb P}}
\def\Prp{{\Bbb P}_{+}}
\def\Prm{{\Bbb P}_{-}}
\def\Prpm{{\Bbb P}_{\pm}}

\def\Prpex{({\Bbb P}_{+})_{\rm exact}}
\def\Prmex{({\Bbb P}_{-})_{\rm exact}}
\def\Prpmex{({\Bbb P}_{\pm})_{\rm exact}}

\def\Pzero{P_0}

\def\Diff{{\Bbb H}}

\def\kp{\kappa_{+}}
\def\km{\kappa_{-}}
\def\kpm{\kappa_{\pm}}

\def\Vlow{E}
\def\Plow{P}


\def\sign{\hbox{\rm sign}}
\def\eps{\epsilon}

\rightline{CERN-TH/2002-350}
\rightline{DESY 02-212}
\rightline{CPT-2002/P.4458}

\vskip 0.8 true cm 
\centerline{\Bigrm Numerical techniques for lattice QCD 
in the $\textfont0=\Bigrm\textfont1=\Bigi\epsilon$--regime}
\vskip 0.6 true cm
\centerline{\bigrm L.~Giusti$^{1,2}$,
C.~Hoelbling$^2$, M.~L\"uscher$^1$, H.~Wittig$^3$}
\vskip1.8ex
\centerline{$^1\hskip-3pt$ \it 
CERN, Theory Division, CH-1211 Geneva 23, Switzerland}
\vskip1.0ex
\centerline{$^2\hskip-3pt$ \it 
Centre Physique Th\'eorique, CNRS, Case 907, Luminy, F-13288 Marseille, France}
\vskip1.0ex
\centerline{$^3\hskip-3pt$ \it 
DESY, Theory Group, Notkestrasse 85, D-22603 Hamburg, Germany}
\vskip 0.8 true cm
\thintablerule
\vskip 2.0ex
\ninepoint
\leftline{\bf Abstract}
\vskip 1.0ex\noindent
In lattice QCD it is possible, in principle, to 
determine the parameters in the effective chiral lagrangian
(including weak interaction couplings)
by performing numerical simulations 
in the $\eps$--regime, i.e.~at quark masses where the 
physical extent of the lattice is much smaller
than the Compton wave length of the pion.
The use of a formulation of the lattice theory that
preserves chiral symmetry is attractive in this context,
but the numerical implementation of any such approach
requires special care in this kinematical situation due to the presence of 
some very low eigenvalues of the Dirac operator.
We discuss a set of techniques
(low-mode preconditioning and adapted-precision algorithms in particular)
that make such computations numerically safe and more efficient
by a large factor.
\vskip 2.0ex
\thintablerule

\tenpoint

\vskip-0.3cm

\vskip-0.1cm

\section 1. Introduction

At low energies the physics of the light pseudo-scalar mesons
can be described by an effective chiral $\sigma$--model with 
$\SUthree_{\rm left}\times\SUthree_{\rm right}$ symmetry group.
Purely strong interaction phenomena and
electroweak transitions (such as the non-leptonic kaon decays)
are both covered by the effective theory if the appropriate
interaction terms are included in the lagrangian.
The challenge is then to compute the associated coupling constants
from the underlying field theory of the strong and the electroweak
interactions.

In the effective chiral theory the physical amplitudes are obtained
in the form of an asymptotic expansion
in powers of the quark masses and the meson momenta.
The coupling constants appear in the coefficients of this 
expansion, and to determine their values 
it is therefore essential
to choose a computational strategy where the limit
of small masses and momenta can be safely reached.

The so-called $\eps$--regime of QCD
[\ref{GasserLeutwylerEpsI}--\ref{LeutwylerSmilga}],
combined with a formulation of lattice QCD
in which chiral symmetry is exactly preserved
[\ref{GinspargWilson}--\ref{Locality}],
provides a framework that may prove to be particularly
suitable in this context.
Some encouraging results have in fact already been obtained
along these lines in the case of the quark condensate
[\ref{EdwardsHellerNarayanan}--\ref{HasenfratzHauswirthEtAl}]
which is one of the free parameters in the chiral lagrangian.
In general numerical simulations in the $\eps$--regime are
technically demanding, however,
because the lattice Dirac operator tends to be ill-conditioned. 
The problem is linked to the 
spontaneous breaking of chiral symmetry 
[\ref{LeutwylerSmilga},\kern1pt%
\ref{ShuryakVerbaarschot}--\ref{DamgaardNishigaki}]
and is hence present independently of how precisely
the lattice theory is defined.

In this paper we discuss a set of fast numerical
methods for the computation of the 
quark propagator and the zero-modes of the Dirac operator.
Being efficient is clearly very important here,
but we wish to emphasize that an ill-conditioned system
also gives rise to questions of numerical stability and accuracy
[\ref{GolubLoan}]
that must be properly dealt with.

Although some of our techniques are more widely applicable,
we focus on the case 
of Neuberger's lattice Dirac operator [\ref{NeubergerDirac}]  
in this paper.
After a brief review of the $\eps$--regime in sect.~2, we then first 
discuss a particular numerical implementation of this operator,
following previous work on the subject 
[\ref{Locality},\kern1pt\ref{Bunk}--\ref{vandenEshofRational}],
which is uniformly accurate to a specified level of precision
(sects.~3--5). 
The ability to rigorously control the 
approximation error is seen to be very useful in sect.~6,
where we describe an efficient way to calculate the index 
of the Dirac operator, and it also provides the basis 
for the adapted-precision inversion algorithm 
discussed in sect.~9. Before this, in sects.~7 and 8, we 
show how the zero-mode contribution to the quark propagator 
can be safely separated
and introduce a method, referred to as {\it low-mode
preconditioning}, that takes care (to some extent)
of the potentially very large condition number
of the Dirac operator in the subspace orthogonal to the zero-modes.

\section 2. The $\epsilon$--regime of QCD 

In numerical simulations of lattice QCD the adjustable parameters
are the gauge coupling, the quark masses and the lattice size.
Usually a euclidean 
$T\times L^3$ lattice is considered, with periodic boundary
conditions, where $T$ and $L$ refer to the time and the space
directions respectively. We shall also assume in this paper
that the lattice Dirac operator satisfies the
Ginsparg--Wilson relation so that 
chiral symmetry is exactly preserved on the lattice
[\ref{ExactChSy}].
The $\eps$--regime is then reached at fixed gauge coupling 
by scaling $T$ and $L$ to values much larger than the confinement 
radius and simultaneously the light quark masses $m=m_u,m_d,m_s$ 
to zero in such a way that the combination
\equation{
  x=m\Sigma V, 
  \qquad
  V\equiv TL^3,
  \enum
}
remains fixed. In this equation the parameter
\equation{
  \Sigma=\lim_{m\to0}\lim_{V\to\infty}\left|\langle\ubar u\rangle\right|
  \enum
}
is the (unrenormalized) $u$--quark condensate in infinite volume.%
\footnote{$\dag$}{\footnotefont%
For simplicity we do not consider the case of quenched or 
partially quenched QCD in this section. Moreover 
we stick to a continuum notation
and drop all factors that formally converge to $1$ in the continuum limit.}

\subsection 2.1 Effective chiral theory

To leading order the purely strong interaction part of the euclidean
action density of the effective theory reads
\equation{
  {\cal L}=\frac{1}{4}F^2\,
  \tr\left\{\partial_{\mu}U^{\dagger}\partial_{\mu}U\right\}
  -\frac{1}{2}\Sigma\,\tr\left\{UM+M^{\dagger}U^{\dagger}\right\},
  \enum
}
where $U$ denotes the SU(3)--valued chiral field and 
$M=\hbox{diag}\left(m_u,m_d,m_s\right)$ the 
quark mass matrix.
The normalizations chosen here are such that the 
parameters $F$ and $\Sigma$ coincide 
with the pion decay constant and the chiral condensate at tree-level
of the chiral perturbation expansion.

In the chiral limit the effective theory is expected to provide a
correct description of the low-energy properties of lattice QCD,
up to lattice spacing effects, if 
$F$ and $\Sigma$ 
are set to the proper values in 
units of the lattice spacing.
These parameters can thus be determined by
calculating any suitable correlation functions on the lattice
at small masses and momenta
and by comparing the results  
with the predictions of the effective theory.
In particular, 
such computations may be performed in the 
$\eps$--regime, where the quark mass term breaks 
chiral symmetry only very weakly 
and where the zero-momentum modes $U(x)=V$
dominate the partition function of the effective theory
[\ref{GasserLeutwylerEpsI}].
There are many different 
ways to study the matching between the effective theory 
and lattice QCD in this regime, most of which
still need to be explored.

An example that we wish to quote for illustration 
is the two-point function of the left-handed current
\equation{
  J^a_{\mu}=\psibar\lambda^a\dirac{\mu}P_{-}\psi,
  \qquad
  P_{\pm}\equiv\frac{1}{2}(1\pm\dirac{5}),
  \enum
}
which is represented by 
\equation{
  \hat{J}^a_{\mu}=\frac{1}{2} F^2\,
  \tr\left\{\lambda^a U\partial_{\mu}U^{\dagger}\right\}
  \enum
}
in the effective theory.
Standard notations are being used here, where $\psi$ denotes the quark field
with flavour components $(u,d,s,\ldots)$ and
$\lambda^a$ an SU(3) generator that acts on the flavour
indices of the fields.
At non-zero time separations and vanishing spatial momentum,
this correlation function turns out to be
proportional to $F^2/T$ 
[\ref{HansenEps},\ref{HansenLeutwyler},\ref{DamgaardHernandezEtAl}]
and it is thus expected to be
particularly suitable for the calculation of the parameter $F$.

\subsection 2.2 Weak-interaction couplings

The effective couplings associated with hadronic weak transitions
can perhaps be computed in a similar way
by matching correlation functions in the $\eps$--regime.
In the case of 
the CP conserving $K\to\pi\pi$ decays, for example, 
the weak interaction density that is obtained after integrating
out the $W$--bosons is,
to a good approximation, equal to
a linear combination of 
[\ref{Weinberg}--\ref{CapitaniGiusti}]
\equation{
  {\cal O}^{\pm}=
  \left\{(\sbar\dirac{\mu}P_{-}u)(\ubar\dirac{\mu}P_{-}d)\pm
         (\sbar\dirac{\mu}P_{-}d)(\ubar\dirac{\mu}P_{-}u)\right\}
  - \left(u\to c\right),
  \enum
  \next{2.0ex}
  {\cal O}_m=\left(m_c^2-m_u^2\right)\left\{
  m_d\left(\sbar P_{+}d\right)+m_s\left(\sbar P_{-}d\right)\right\}.
  \enum
}
The operator ${\cal O}_m$
is usually omitted from this list, because it
does not contribute in transition matrix elements
where the total energy-momentum is conserved.
Mixings of ${\cal O}^{\pm}$ and ${\cal O}_m$ with other operators are, 
incidentally, excluded 
by the exact sym\-me\-tries of the lattice theory.

In the effective chiral theory, the operators ${\cal O}^{\pm}$
and ${\cal O}_m$
are expected to be represented by certain linear combinations of 
the fields [\ref{BernardEtAl}]
\equation{
  \Ohat_1=\frac{1}{8}F^4\bigl\{
  (U\partial_{\mu}U^{\dagger})_{13}
  (U\partial_{\mu}U^{\dagger})_{21}+
  \noenum
  \next{1.5ex}
  \kern5.1em
  (U\partial_{\mu}U^{\dagger})_{23}
  (U\partial_{\mu}U^{\dagger})_{11}
  +\frac{1}{5}(\partial_{\mu}U\partial_{\mu}U^{\dagger})_{23}
  \bigr\},
  \enum
  \next{2.0ex}
  \Ohat_2=
  \frac{1}{8}F^4\kern1pt(\partial_{\mu}U\partial_{\mu}U^{\dagger})_{23},
  \enum
  \next{2.0ex}
  \Ohat_3=
  \frac{1}{2}F^2\Sigma\kern1pt
  (UM+M^{\dagger}U^{\dagger})_{23},
  \enum
}
and our task is then to determine the dimensionless coefficients
in these linear combinations. 
An obvious possibility
in the $\eps$--regime is to match three-point correlation functions such as
\equation{
  C^{ab}(x_0,y_0)=\int_0^L\rmd^3x\kern1pt\rmd^3y\kern1pt\left\langle 
  J^a_{\mu}(x){\cal O}^{\pm}(0) J^b_{\mu}(y)\right\rangle
  \enum
}
plus a similar correlation function with only one left-handed current
(so as to be able to disentangle the contributions of 
the octet operators $\Ohat_2$ and $\Ohat_3$).
Explicit computations in chiral
perturbation theory suggest that these correlation functions 
are indeed well-behaved 
and that, in principle, the desired coefficients can be 
calculated along these lines [\ref{PilarMikko}].

\subsection 2.3 Spectrum of the Dirac operator

The last ten years or so have seen a remarkable development,
involving both random matrix and chiral perturbation theory,
that culminated in the analytical determination of the
distributions of the low-lying eigenvalues
of the Dirac operator in the $\eps$--regime
[\ref{LeutwylerSmilga},\ref{ShuryakVerbaarschot}--\ref{DamgaardNishigaki}]
(for a review see ref.~[\ref{DamgaardReview}]).
An important insight gained in the course of this work is
that the eigenvalues scale like $(\Sigma V)^{-1}$ at 
large volumes. In particular, the spectral gap
in the subspace orthogonal to the chiral zero-modes
is, on average, roughly of this size.

For the calculation of the quark propagator in the $\eps$--regime,
the presence of some very low 
eigenvalues (an effect that has been confirmed numerically on small 
lattices
[\ref{EdwardsHellerKiskisNarayanan},\ref{DamgaardEdwardsHellerNarayanan},%
\ref{HasenfratzHauswirthEtAl}])
is a source of difficulty.
To make this a bit more concrete,
let us consider a $32^4$ lattice with spacing $a=0.1\,\fm$
and let us assume that $\Sigma$ is around $(250\,\MeV)^3$
on this lattice.
From the discussion above we then deduce that 
the expectation value of the spectral gap
is about $5\times10^{-4}$ in lattice units.
Note that the quark masses
are scaled proportionally 
to $(\Sigma V)^{-1}$ in the $\eps$--regime and thus do not
provide a strong infrared cutoff on the spectrum.

The computation of the quark propagator, using
the conjugate gradient algorithm for example, is perhaps still possible
under these conditions, but may require a very large
number of iterations. Moreover,
the accuracy of the solution vector that can be reached
on any given machine may be unsatisfactory, especially
if the propagator is calculated via the
so-called normal equations 
(as is the case if the conjugate gradient algorithm is used) 
[\ref{GolubLoan}].
Effectively these technical problems set a limit on the 
space-time volumes where numerical simulations can be performed
and thus on how close to the chiral point one can get in practice.

\section 3. Lattice Dirac operator

We set up the lattice theory as usual on
a finite four-dimensional lattice with spacing $a$ and 
periodic boundary conditions. 
Our notational conventions are summarized in appendix A.
Following refs.~[\ref{GinspargWilson}--\ref{ExactChSy}]
we consider a formulation of lattice QCD where 
the lattice Dirac operator $D$ satisfies the Ginsparg--Wilson relation
\equation{
   \dirac{5}D+D\dirac{5}=\abar D\dirac{5}D,
   \qquad
   D^{\dagger}=\dirac{5}D\dirac{5},
   \enum
}
for some positive constant $\abar$ proportional to $a$.
In the presence of a given background gauge field,
the propagator of a quark with mass $m$ is then given by
\equation{
  \left\langle\psi(x)\psibar(y)\right\rangle=
  \bigl\{D_m^{-1}\bigr\}(x,y),
  \enum
}
where $\{\ldots\}(x,y)$ stands for the kernel in position space 
of the operator in the curly bracket and 
\equation{
  D_m=(1-\frac{1}{2}\abar m)D+m
  \enum
}
is the massive lattice Dirac operator. We shall always assume that
$0\leq\abar m\leq2$.

As already mentioned in sect.~1, 
we concentrate on the case of
the Neuberger--Dirac operator in this paper,
even though many of the techniques that will be described 
are more generally applicable.
Explicitly this operator is given by [\ref{NeubergerDirac}] 
\equation{
  D={1\over\abar}\left\{1+\dirac{5}\,\sign(Q)\right\},
  \enum
  \next{2ex}
  Q=\dirac{5}\left(a\Dw-1-s\right),
  \qquad
  \abar={a\over1+s},
  \enum
}
where $s$ is an adjustable parameter in the range $|s|<1$
and $\Dw$ denotes the standard Wilson--Dirac operator (appendix A).

\section 4. Minmax polynomial approximation

For the numerical implementation of 
the Neuberger--Dirac operator 
we need to find an algorithm 
that evaluates
$\sign(Q)\eta$ on any given fermion field $\eta$
to a specified precision.
In the present section we take a first step in this direction by
constructing an optimal polynomial approximation to $\sign(x)$ in 
a range that excludes a small interval around the origin.
More precisely we are looking for a polynomial $P(y)$ of degree $n$
that minimizes the error
\equation{
  \delta=\max_{\eps\leq y\leq 1}\left|h(y)\right|,
  \qquad
  h(y)\equiv1-\sqrt{y}P(y),
  \enum
}
for specified $\eps>0$. 
In the range $\sqrt{\eps}\leq |x|\leq1$ the function 
$xP(x^2)$ then approximates $\sign(x)$ uniformly with a maximal
deviation $\delta$.

\subsection 4.1 Existence and uniqueness

Polynomial approximations that minimize the maximal relative
error are referred to as minmax polynomials.
They can be shown to exist and to be uniquely determined
provided the function that is to be approximated does not vanish.
Moreover, practical methods have been devised, based on 
the characteristic properties of the associated error function,
that allow one to compute them reliably
(see ref.~[\ref{Handscomb}], for example, 
or any other good book on approximation theory).

For reasons of numerical stability, 
it is advantageous to represent the polynomial
in the form of a series 
\equation{
  P(y)=\sum_{k=0}^nc_kT_k(z),
  \qquad
  z=(2y-1-\eps)/(1-\eps),
  \enum
}
of Chebyshev polynomials (appendix A).
Our task is then to adjust the coefficients 
$c_k$ so that the deviation (4.1) is minimized.

\subsection 4.2 Error bounds

The characteristic property alluded to above of the minmax polynomial 
is that the error function $h(y)$ has (at least) 
$n+2$ extrema of equal height and alternating sign, 
i.e.~$h(y)$ is oscillating around zero with constant amplitude
(fig.~1).
Evidently it is not possible to know in advance where these extrema
are located.

Let us now choose an arbitrary sequence
\equation{
  \eps\leq y_0<y_1<\ldots<y_{n+1}\leq1
  \enum
}
of $n+2$ values.
We can then
find a polynomial such that
\equation{
  h(y_l)=(-1)^lu\quad\hbox{for all}\quad l=0,\ldots,n+1.
  \enum
}
In fact these are just $n+2$ linear equations
in the unknowns $c_0,\ldots,c_n,u$. This polynomial
has the alternating property on the chosen set of points,
but not necessarily so on the approximation interval.
It can be shown, however, that
\equation{
  \left|u\right|
  \leq 
  \max_{\eps\leq y\leq1}\left|h_*(y)\right|
  \leq
  \max_{\eps\leq y\leq1}\left|h(y)\right|
  \enum
}
where $h_*(y)$ denotes the error function of the minmax polynomial.
The solution of the linear equations (4.4) thus yields a nearly optimal 
polynomial if these bounds are tight.

\topinsert
\vbox{
\vskip0.0cm
\epsfxsize=8.0cm\hskip2.0cm\epsfbox{figure1.eps}
\vskip0.4cm
\figurecaption{%
Minmax polynomial approximation of sign$(x)$ for $\eps=0.0025$
and $n=22$ (full line). 
In the range $\sqrt{\eps}\leq |x|\leq1$
the relative error of this approximation is $5\%$,
while for an error of $0.1\%$ a polynomial of degree $n=57$ 
is required (dashed line).
}
\vskip0.0cm
}
\endinsert

\subsection 4.3 Maehly's exchange algorithm

The basic idea of this algorithm is to start from some well-chosen 
initial set of points
(4.3) and to adjust them subsequently so as to tighten the bounds (4.5).
This can be achieved, for example, by choosing the new points to be the 
nearby extremal points of the current error function. 

A good initial set of points for this recursion are the extremal points 
of the Chebyshev polynomial $T_{n+1}(z)$.
Only few iterations are then required until a polynomial
is obtained that is very nearly optimal. There is in fact little to be
gained by performing further iterations once the lower and the upper
bound in (4.5) differ by no more than (say) $0.01\left|u\right|$.
Evidently this procedure is extremely robust,
because the quality of the current polynomial (with respect to the best
possible approximation) can be rigorously controlled.

\subsection 4.4 Implementation details

Inspection shows that the linear system (4.4) is well-conditioned
and it can thus be solved straightforwardly by $QR$ factorization, using
Householder transformations, followed by backward substitution
(see ref.~[\ref{GolubLoan}] for example). 

To localize the extrema of $h(y)$,
a few bisection steps are sufficient in general,
since there is no need to be very precise at this point.
An important detail here is that
many digits are lost when calculating the error function, 
even if the Clenshaw recursion is used
to evaluate the Chebyshev series (4.2) [\ref{Recipes}].
This becomes a problem 
when $\delta/12n$ approaches the machine precision, and
with standard 64 bit arithmetic
the approximation accuracy $\delta$ 
that can be reached is thus limited to values
greater than $n\times10^{-15}$ or so.

\subsection 4.5 Efficiency of the approximation

The precision of the approximation to the sign function
provided by the minmax polynomial scales approximately like
\equation{
  \delta=A\rme^{-bn\sqrt{\eps}}
  \enum
}
where $A=0.41$ and $b=2.1$.
At large values of $n\sqrt{\eps}$,
this empirical formula is fairly accurate,
but it
slightly over-estimates the actual value of $\delta$
when the degree of the polynomial is less than 30 or so.
With respect to the Chebyshev approximation
that was used in ref.~[\ref{Locality}], 
the precision is improved by about a factor of $2$.

We finally mention that, on a current PC processor and 
for degrees $n\leq400$,
only a few seconds are required 
to obtain the coefficients $c_k$ of the minmax polynomial.
It is hence practical to recompute these polynomials 
whenever they are needed.

\section 5. Uniform approximation of the Neuberger--Dirac operator

The minmax polynomial $P_{n,\eps}(y)$ of degree $n$ 
on the interval $[\eps,1]$ 
yields an approximation of the operator $\sign(Q)$ through
\equation{
   \sign(Q)\simeq XP_{n,\eps}(X^2),
   \qquad
   X\equiv Q/\|Q\|.
   \enum
}
If $\eps$ is chosen so that $Q^2\geq\eps\|Q\|^2$,
the error in this formula
is an operator with norm less than or equal to $\delta$.
In other words, 
the approximation error is always bounded by $\delta\|\eta\|$,
uniformly in the field $\eta$ to which the operator
is applied.

This method is straightforward but actually not recommendable
if $Q^2$ has some exceptionally low eigenvalues
(as is often the case in practice [\ref{Locality}]).
It is far more efficient under these conditions to first
separate the few lowest modes and to treat them exactly.
Evidently this should be done in such a way that the 
total approximation error remains under control.

\subsection 5.1 Low-mode projectors

The spectrum of $Q$ in the vicinity of the origin
can be reliably determined by mini\-mizing the Ritz functional 
of $Q^2$ [\ref{EvaI},\ref{EvaII}].
This technique also yields an approximation to the associated eigenvectors,
and we now need to discuss by how much these 
vectors deviate from the true eigenvectors. 

So let us assume that 
a specified number $l$ of approximate eigenvectors has been
computed.
We denote by $V$ the linear space spanned by 
these vectors and by $\Pr$ the corresponding orthonormal projector.
It is then trivial to calculate the parameter
\equation{
  \varrho=\max_{v\in V, \|v\|=1}\left\|(1-\Pr)Qv\right\|
  \enum
}
which measures the deviation of $V$ from being an exact eigenspace of $Q$.
The importance of this parameter is clarified by

\proclaim Lemma 5.1.
Let $\nu_1,\ldots,\nu_l$ be the eigenvalues of\/ $\Pr Q\Pr$ in
the subspace $V$. Then there are $l$ linearly independent
eigenvectors of $Q$ with
eigenvalues $\lambda_1,\ldots,\lambda_l$ such that
$|\nu_k-\lambda_k|\leq\varrho$ for all $k=1,\ldots,l$.

\noindent
The proof of the lemma (which holds independently of how $V$ was obtained)
is given in appendix B.

In the following we shall take it for granted
that the eigenvalues $\nu_k$ are separated from zero 
and from the rest of the spectrum of $Q$ (all eigenvalues $\lambda$ except
$\lambda_1,\ldots,\lambda_l$)
by a distance greater than $\varrho$.
So far we have in fact never encountered a situation where this condition
could not be satisfied, and we thus omit any further discussion of this point.
The presence of a spectral gap 
around zero implies that the subsets of 
positive and negative eigenvalues
can be identified without any numerical ambiguity.
If we introduce the eigenvectors
\equation{
  u_k\in V,\qquad
  \Pr Qu_k=\nu_ku_k, 
  \qquad
  (u_k,u_j)=\delta_{kj},
  \enum
}
the associated orthonormal projectors are given by
\equation{
  \Prp=\sum_{\nu_k>0} u_k\otimes(u_k)^{\dagger}
  \quad\hbox{and}\quad
  \Prm=\sum_{\nu_k<0} u_k\otimes(u_k)^{\dagger}
  \enum
}
respectively. In the same way we may define the projectors $\Prpmex$
to the subspaces spanned by the corresponding 
exact eigenvectors of $Q$
(whose existence is guaranteed by lemma 5.1)
with positive and negative eigenvalues $\lambda_k$. 

We may now ask how accurately the computed
projectors $\Prpm$ approximate the exact projectors $\Prpmex$.
The answer to this question depends on the size of 
the residues
\equation{
 \varrho_k=\left\|(Q-\nu_k)u_k\right\|
 \enum
}
and also on the distances between the eigenvalues of $Q$.
A small value of $\varrho_k$ alone does, in fact, not 
exclude sizeable mixings of $u_k$ with several eigenvectors
of $Q$ if these have eigenvalues that 
are within a distance $\varrho_k$ of $\nu_k$.

Eventually we are interested in estimating the 
deviation of the projectors
rather than that of the individual eigenvectors, and
what counts in this case is the distance $d_k$ of $\nu_k$ from 
the exact spectrum of $Q$ in the subspace
that is orthogonal to the range of $\Prpex$ if $\nu_k>0$
or $\Prmex$ if $\nu_k<0$. The quality of the approximation 
is then controlled by the parameters
\equation{
  \kpm^2=
  \sum_{\pm\nu_k>0}\varrho_k^2/d_k^2,
  \qquad \kpm>0,
  \enum
}
as the following lemma shows.

\proclaim Lemma 5.2.
Provided the spectral conditions listed above are satisfied, the bound
\equation{
  \|\Prpm-\Prpmex\|\leq{\kpm
  (1+2\kpm)\over1-2\kpm(1+2\kpm)}
  \enum
}
holds where it is assumed that\/ $2(l+1)\kpm(1+2\kpm)<1$.

\noindent
The proof of the lemma is given in appendix C. Here we only note
that the parameters $\kpm$ are very small in the cases of interest.
The right-hand side of eq.~(5.7) is 
then practically equal to $\kpm$ and the condition on the 
last line is trivially fulfilled.

\subsection 5.2 Approximation formula and error bound

It is now straightforward
to write a program that computes
the projectors to a specified accuracy. 
In this program the number of low modes
to be included in the projectors 
should be determined dynamically in such a way that
the spectral distance from the other modes is not accidentally 
very small. 
The parameters $\kpm$ can then be estimated
without difficulty and the minimization of the Ritz functional
is stopped when the desired level of precision is reached.

Once the projectors are computed, eq.~(5.1) may be replaced by
\equation{
  \sign(Q)\simeq\Prp-\Prm+\left(1-\Prp-\Prm\right)X
  P_{n,\eps}(X^2),
  \qquad
   X\equiv Q/\|Q\|,
  \enum
}
where $\eps$ is set to a value equal to (or perhaps slightly less than)
the smallest eigenvalue of $Q^2/\|Q\|^2$ in the sector orthogonal to 
the low modes. 
The properties of 
the minmax polynomial, lemma 5.2 and simple
triangle inequalities then imply that the associated 
approximation $\widetilde{D}_m$ to the massive
Neuberger--Dirac operator satisfies
\equation{
  \|\widetilde{D}_m-D_m\|\leq
  {1\over a}\,(1+s-\frac{1}{2}am)\left\{2(\kp+\km)+\delta\right\}
  \enum
}
up to terms proportional to $\kpm\delta$ and $\kpm^2$
(from now on these will be neglected).

\subsection 5.3 Miscellaneous remarks

In the tests that we have performed we never had any difficulty 
in reaching accuracy levels $\kpm$ of $10^{-8}$ or even $10^{-10}$.
Moreover, the actual approximation error is
significantly smaller than suggested by lemma 5.2, 
by a factor $10$ at least.
Since the calculation of the projectors consumes a relatively small
amount of computer time, they will normally be obtained to 
a precision that is estimated to be sufficient, by a wide margin,
for the applications that one has in mind. The approximation error
(5.9) is then essentially determined by the degree $n$ of the minmax
polynomial.

A numerically safe method to evaluate 
the polynomial in eq.~(5.8) is provided by
the Clenshaw recursion [\ref{Recipes}].
The accumulated rounding errors are then on the
order of $n$ times the machine precision.
In particular,
if standard 32 bit floating-point arithmetic is used
(which can be profitable at intermediate
stages of the algorithms described later),
a rounding error of about $5n\times10^{-7}$ should be added to
the theoretical approximation error $\delta$.

We finally note that the degree of the minmax polynomial
will usually be significantly larger than the number of the separated
low modes. The application of the projectors in eq.~(5.8)
then requires only a small fraction of the total execution time
and the associated rounding errors can also be safely neglected.

\section 6. Calculation of the index of the Dirac operator

\vskip-1.5ex

\subsection 6.1 Preliminaries

The operator $D^{\dagger}D$
commutes with $\dirac{5}$ and thus leaves the subspaces of fermion
fields with definite chirality invariant.
Using the Ginsparg--Wilson relation, it can be shown
that the action of the operator in these subspaces is given by the
hermitian operators
\equation{
  \Dzpm=P_{\pm}DP_{\pm}
  \enum
}
up to a proportionality factor equal to $2/\abar$.
Moreover, the spectrum of the low-lying eigenvalues of $\Dzp$ and $\Dzm$
is exactly the same, including degeneracies,
apart from the fact that there can be a surplus of zero-modes in
one of the sectors. 

It happens only accidentally (with probability zero strictly speaking)
that there are zero-modes with both chiralities,
and we shall thus exclude this exceptional case in the following.
The index $\nu$ of the Dirac operator is then given by
\equation{
  \nu=\sigma n_0
  \enum
}
where $\sigma$ denotes the chirality of the sector that 
contains the zero-modes and $n_0$ the number of these modes.

\subsection 6.2 Zero-mode counting

The low-lying eigenvalues of $\Dzp$ and $\Dzm$
can, in principle, be found by minimizing the associated Ritz functionals
[\ref{EvaI},\ref{EvaII}].
In terms of computer time such calculations tend to be very expensive,
however, since each application of the Neuberger--Dirac operator
to a given fermion field
involves from say 50 to several 100 applications of $Q^2$.
Any numerical method that requires a
precise computation of the eigenvalues
should hence be avoided.

The strategy that we propose is to run the minimization program
in both chirality sectors simultaneously in a controlled way,
where the precision is improved in steps by a specified
factor.
After every iteration the current estimate of the 
lowest eigenvalue in each sector provides a rigorous upper bound
on the gap above zero in the spectrum of $\Dzp$ and $\Dzm$ respectively.
The program then proceeds to lower the larger bound
and continues in this manner until the estimated relative error on any
one of the eigenvalues in the two sectors is determined to be less than $10\%$. 
At this point we know that 
the zero-modes (if any) must be in the other chirality sector.

To count the zero-modes 
the minimization program must now be run once more in that sector,
improving the precision on the calculated eigenvalue in 
steps as before. As soon as the eigenvalue drops below
the spectral gap in the other sector (which was determined
to an accuracy of $10\%$ in the previous calculation), 
there has to be at least one zero-mode, 
independently of the current level of precision,
because the Ritz functional always provides 
an upper bound on the true eigenvalue.
We can then restart the minimization program, requiring a second
eigenvalue to be computed, and so on. 

This procedure terminates when an eigenvalue is found that
is definitely above zero, which will be the case if the 
relative error
is estimated to be less than $10\%$ for example.
It should be emphasized that in this way we never 
require any eigenvalues to be calculated to high precision
and yet the zero-mode counting is rigorously correct.

\subsection 6.3 Using reduced-precision operators

In the discussion above we have implicitly assumed that 
the approximation error (5.9) of 
the numerical representation $\Dtildepm$ of the operators $\Dzpm$ 
can be neglected. An important point to note is 
that the associated error on the eigenvalues is at most
\equation{
   \omega={1\over a}\,(1+s)\left\{2(\kp+\km)+\delta\right\},
   \enum
}
and the same is true for the gradients of the 
Ritz functionals of $\Dzpm$. We may, therefore,
start the minimization of the Ritz functionals 
using a relatively poor approximation of the Dirac operator,
and then gradually decrease $\delta$ so that the error 
$\omega$ is always smaller than
the current magnitude of the gradients. 

Apart from being in control of the numerical errors,
an obvious advantage of this procedure is that
the degree of the minmax polynomial is not larger than absolutely
necessary, at all stages of the calculation.
The total computational effort is thus minimized.
Moreover, a further acceleration
may be achieved by using single-precision arithmetic
in the programs for the Neuberger--Dirac operator
as long as $\delta$ is not too small (cf.~subsect.~5.3).
On current PC processors this saves almost a factor of $2$
in execution time [\ref{SSE}].

\subsection 6.4 Error estimation

In the algorithm described in subsect.~6.2, the error on the 
calculated eigenvalues must be estimated in a reliable way.
For any given eigenvalue,
the error is a sum of two contributions, one coming from the 
numerical approximation of the lattice Dirac operator
and the other from the fact that, in each step of
the procedure, the minimum of the relevant Ritz functional is only found to 
some specified accuracy.

A reliable but often poor estimate of the latter error
is provided by the gradient of the Ritz functional
[\ref{EvaI},\ref{EvaII}]. However, for reasons of numerical stability
it is in any case advisable to compute an additional
eigenvalue in each sector, 
with possibly reduced precision [\ref{EvaII}],
and the Temple inequality
may then be used to obtain a more accurate estimate 
of the error on the other eigenvalues [\ref{ReedSimon}].

\section 7. The quark propagator: general strategies 

The presence of chiral zero-modes 
(which is the normal case in large volumes)
complicates the calculation of 
the quark propagator, and it is mainly this issue
that we wish to address in this section.
We assume in the following that the quark mass $m$ is strictly positive
and that a gauge field configuration is being considered where
the zero-modes (if any) have chirality $\sigma=+1$.

\subsection 7.1 Negative chirality components

The computation of the quark propagator amounts to solving the 
linear equation
\equation{
  \Dm\psi=\eta
  \enum
}
for a set of source fields $\eta$. We first write the solution in the 
form
\equation{
  \psi=(\Dmdag\Dm)^{-1}\Dmdag\eta
  \enum
}
and note that the operator in brackets commutes with $\dirac{5}$. 
The components of the solution vector with negative chirality 
are thus given by
\equation{
  \Pm\psi=(\Dmdag\Dm)^{-1}\Pm\Dmdag\eta.
  \enum
}
In this equation the inversion of $\Dmdag\Dm$ takes place in the 
chirality sector that does not contain the zero-modes, and 
the only potential difficulty is then that there can be some
very low non-zero eigenvalues of $D^{\dagger}D$ in this subspace
(cf.~subsect.~2.4).

We shall come back to this problem
in the next two sections, where we discuss the inversion of 
$\Dmdag\Dm$ in the negative chirality sector.
For the time being, we merely assume that an efficient 
algorithm is available which allows us to invert the operator
in this sector, to a specified precision and 
at arbitrarily small quark masses.

\subsection 7.2 Positive chirality components

Once the negative chirality component
$\Pm\psi$ has been computed, it can be inserted on the 
right-hand side of the equation
\equation{
  \Dm\Pp\psi=\eta-\Dm\Pm\psi
  \enum
}
for the positive chirality components $\Pp\psi$.
A possible expression for the solution of this equation is
\equation{
  \Pp\psi=(\Pp\Dm\Pp)^{-1}\left\{\Pp\eta-\Pp\Dm\Pm\psi\right\}
  \enum
}
which involves an inversion of $\Pp\Dm\Pp$ in the
sector that contains the zero-modes.

It is now very important to note that
\equation{
  \Pp\Dm\Pp=\Pp\left\{
  \frac{1}{2}\abar\left(1-\frac{1}{2}\abar m\right)D^{\dagger}D+m
  \right\}\Pp
  \enum
}
is a positive operator in this sector whose condition number
is at most $2/\abar m$ and thus comparatively small.
If the quark mass in lattice units 
is not too close to zero, the inversion in eq.~(7.5) is
hence well-conditioned and 
the solution can be computed straightforwardly, using
the conjugate gradient algorithm for example.

\subsection 7.3 Separation of the zero-modes

Arbitrarily small quark masses can also be handled 
but require a subtraction of
the zero-mode component from the source field.
Equation (7.5) then becomes
\equation{
  \Pp\psi={1\over m}\Pzero\Pp\eta+
  (\Pp\Dm\Pp)^{-1}\left\{
  (1-\Pzero)\Pp\eta-\Pp\Dm\Pm\psi\right\},
  \enum
}
where $\Pzero$ denotes the projector to the subspace spanned by 
the zero-modes.
Note that the field in the curly bracket is orthogonal 
to this subspace, and the operator inversion in eq.~(7.7) thus remains
well-defined even in the limit of a vanishing quark mass.
The situation is then practically the same as in
the negative chirality sector discussed in subsect.~7.1. 
In particular, the acceleration
techniques described in the next two sections can be applied 
here too.

At this point we still need to explain how to compute the 
zero-modes (and thus the 
projector $\Pzero$).
First recall that a basis of approximate
zero-modes is obtained
in the course of the 
calculation of the index of the Dirac operator (cf.~subsect.~6.2).
This basis may not be
very accurate, but we can now easily improve on the accuracy 
by noting that 
\equation{
  \Pzero\chi=\left(1-\frac{1}{2}\abar D\right)\Pp\chi
  -D(D^{\dagger}D)^{-1}\Pm D^{\dagger}\Pp\chi
  \enum
}
for any field $\chi$. The operator inversion in this formula
is in the sector with negative chirality 
(where there are no zero-modes) which 
we have assumed to be feasible. 

It is our experience that
this procedure is numerically stable and yields 
accurate results when applied to the 
approximate zero-modes. More precisely, the 
error on the calculated zero-modes [after the application of eq.~(7.8)]
is estimated to be at most
$r+2\omega/|\lambda_1|$ in this case,
where $\omega$ is the accuracy of the approximation to
the Neuberger--Dirac operator used in the numerator,
$\lambda_1$ the first non-zero eigenvalue of $D$ and
$r$ the relative residuum associated with the numerical 
inversion of $D^{\dagger}D$.%
\footnote{$\dag$}{\footnotefont%
An argumentation similar to the one at the
beginning of appendix C shows that
the deviation $\|(1-P_0)\chi\|$ of any field
$\chi$ from being an exact zero-mode
is less than or equal to $\|D\chi\|/|\lambda_1|$.
The approximation error can thus be rigorously controlled.}

\subsection 7.4 Left-handed propagator

The correlation functions that we have proposed in sect.~2
to probe the weak-inter\-action operators 
involve only left-handed quark and antiquark fields.
To evaluate these correlation functions,
it thus suffices to compute the purely left-handed
components of the quark propagator, which are given by
\equation{
  \left\{\Pm(\Dmdag\Dm)^{-1}\Pm\Dmdag\Pp\right\}(x,y).
  \enum
}
From this expression it is immediate that the complications
arising from the zero-modes are absent in this case.
In fact we only need to invert $\Dmdag\Dm$ in the 
chirality sector where there are no zero-modes.
Note that if the zero-modes had the opposite chirality,
we would write the propagator in the form
\equation{
  \left\{\Pm\Dmdag\Pp(\Dmdag\Dm)^{-1}\Pp\right\}(x,y)
  \enum
}
so that the 
inversion of $\Dmdag\Dm$ is again in the ``good" sector.

Such correlation functions 
thus appear to be particularly attractive
from the numerical point of view.
Moreover,
their behaviour in the chiral limit tends to be less singular
than is generally the case
[\ref{DiamantiniEtAl},\kern1pt\ref{DamgaardHernandezEtAl},%
\kern1pt\ref{PilarMikko}].

\section 8. Low-mode preconditioning

We now proceed to discuss the linear system
\equation{
  A\psi=\eta, 
  \qquad A\equiv\Dmdag\Dm,
  \enum
}
in the negative chirality sector, assuming as before 
that the zero-modes (if any) are in the other sector.

As already emphasized in sect.~2, 
the presence of some extremely low eigenvalues of $A$
forbids a straightforward application of the established 
inversion algorithms to solve 
the equation. To reduce the condition number of the system,
an obvious possibility is to calculate the few lowest eigenvalues of $A$
and to compute the components of the solution vector 
along the associated eigenspace and its orthogonal complement separately.
However, on large lattices this proposition is
not practical, because the computation of the 
eigenvectors to the required level of precision
is far too expensive in terms of computer time.

In this section we describe a version of 
low-mode preconditioning that works out even if 
the eigenvectors cannot be calculated very accurately.
Effectively the algo\-rithm com\-pen\-sates
for the approximation errors
through a simple block diago\-na\-lization
that can be implemented exactly.

\subsection 8.1 Approximate eigenvectors

So let us assume that $e_1,\ldots,e_n$ is a set of orthonormal 
Dirac fields with negative chirality that satisfy
\equation{
  Ae_k=\alpha_ke_k+r_k,
  \qquad
  \left(e_l,r_k\right)=0,
  \qquad
  \|r_k\|\leq\omega_k\alpha_k,
  \enum
}
for all $k,l$ and some positive numbers $\alpha_k$ and $\omega_k$. 
Evidently, if the bounds $\omega_k$ 
on the residues $r_k$ are small, these fields can be regarded
as approximate eigenvectors of $A$ with eigenvalues $\alpha_k$.
In the following we shall also take it for granted that
\equation{
  (v,Av)\geq\gamma\left\|v\right\|^2,
  \qquad
  \gamma\equiv\max_k\alpha_k,
  \enum
}
for all negative chirality vectors $v$
in the orthogonal complement of the vector space
$\Vlow$ spanned by $e_1,\ldots,e_n$.
This guarantees that $e_1,\ldots,e_n$
approximate the lowest $n$ eigenvectors of $A$
rather than an arbitrary set of eigenvectors.

It should be emphasized that no further 
assumptions need to be made and that the 
error bounds $\omega_k$, in particular,
do not need to be extremely small
(in most cases values of $0.1$ or so will do).
In practice the vectors $e_1,\ldots,e_n$ can be computed
by minimization of the Ritz functional of $A$ and stopping
the program when the estimated relative errors of the calculated
eigenvalues (as determined from the gradient of the Ritz functional 
[\ref{EvaI},\ref{EvaII}]) drop below  
the specified values of $\omega_k$.%
\footnote{$\dag$}{\footnotefont%
Since the eigenvectors of $A$ are mass-independent,
we may set $m=0$ in this calculation and use
the same vectors $e_1,\ldots,e_n$ at all values of $m$.
Only the eigenvalues $\alpha_k$ and the residual vectors $r_k$ 
then need to be recalculated if the quark mass changes. 
}

\subsection 8.2 Preconditioned system

We now introduce the projector $\Plow$
to the subspace $\Vlow$ and
note that the linear system (8.1) is equivalent to
\equation{
  (1-\Plow)\left\{A-A\Plow(\Plow A\Plow)^{-1}\Plow A\right\}(1-\Plow)\psi
  \noenum
  \next{1ex}
  \kern3.0cm=(1-\Plow)\eta-(1-\Plow)A\Plow(\Plow A\Plow)^{-1}\Plow\eta,
  \enum
  \next{2.5ex}
  \Plow A\Plow\psi=\Plow\eta-\Plow A(1-\Plow)\psi.
  \enum
}
The first of these equations determines the component
$(1-\Plow)\psi$ of the solution in the subspace orthogonal to $\Vlow$.
Once it has been calculated, the parallel component, $\Plow\psi$, 
is obtained from the second equation.

To make this a little more explicit, it is helpful to introduce
the field
\equation{
  \phi\equiv(1-\Plow)\psi=\psi-\sum_{k=1}^ne_k\left(e_k,\psi\right).
  \enum
}
Equation (8.4) may then be written as
\equation{
  (1-\Plow)A\phi-\sum_{k=1}^nr_k{1\over\alpha_k}\left(r_k,\phi\right)
  =(1-\Plow)\eta-\sum_{k=1}^nr_k{1\over\alpha_k}\left(e_k,\eta\right).
  \enum
}
Together with the constraint $\Plow\phi=0$, this is a well-defined system
whose solution immediately yields the complete field through 
\equation{
  \psi=\phi+\sum_{k=1}^ne_k{1\over\alpha_k}\left\{\left(e_k,\eta\right)
  -\left(r_k,\phi\right)\right\}.
  \enum
}
The scalar products in these formulae give rise to 
a computational overhead that is completely negligible compared to the 
effort required for the application of the operator $A$. This assumes,
of course, that the decomposition (8.2) has been worked out
before the conjugate gradient program is started and that 
it is possible to keep the vectors $e_k$ and $r_k$ in memory during
the whole calculation.

\subsection 8.3 Condition numbers

The operator on the left-hand side of eq.~(8.7)
is the difference of two operators,
\equation{
   (1-\Plow)A(1-\Plow)-
   \sum_{k=1}^n{1\over\alpha_k}\,r_k\otimes(r_k)^{\dagger},
   \enum
}
that act in the orthogonal complement of $\Vlow$ and that are both 
positive.
The lowest eigenvalue of the first operator is greater or equal to $\gamma$,
while the norm of the second is bounded by
$\sum_{k=1}^n(\omega_k)^2\alpha_k$. This implies that the 
condition number of the preconditioned system (8.7) is at most
\equation{
   (4/\abar^2)\times
   \left\{\gamma-\hbox{$\sum_{k=1}^n(\omega_k)^2\alpha_k$}\right\}^{-1},
   \enum
}
which is practically equal to $4/\abar^2\gamma$ if the error bounds
$\omega_k$ are smaller than $0.2\times n^{-1/2}$ for example.
Since the condition number of the 
original system is about $4/\abar^2$ divided by the lowest eigenvalue
of $A$, a reduction in the condition number by roughly
a factor of $\max_k\alpha_k/\min_k\alpha_k$ is thus achieved.

Evidently this ratio
depends on the gauge field configuration and on
the number of modes that are being subtracted.
With $n$ as small as $4$ we have found
factors of $30$ and more.
An interesting remark in this context is
that the improvement factor is roughly independent of the 
lattice parameters, because the low-lying eigenvalues of $A$ scale
proportionally to $(\Sigma V)^{-2}$ (cf.~subsect.~2.3).
The probability distribution of the 
ratio of the $n$th to the first eigenvalue is in fact 
universal in random matrix theory and analytically computable.

\subsection 8.4 The low-mode contribution --- a precision issue

The program that minimizes the Ritz functional of $A$
initially yields a set of approximate eigenvectors
$\tilde{e}_1,\ldots,\tilde{e}_n$ that 
are orthonormal to machine precision but that are not guaranteed to
satisfy eq.~(8.2) very accurately.
We can, however, easily correct for this 
deficit by defining the vectors $e_k$ to be
the exact eigenvectors of $A$ in the subspace spanned by the 
given vectors.
The numbers $\alpha_k$ are then the eigenvalues of the matrix
\equation{
   M_{kl}=(\tilde{e}_k,A\tilde{e}_l)
   \enum
}
and the residues $r_k$ are obtained through
$r_k=(1-\Plow)Ae_k$.

While the subspace spanned by the eigenvectors is
by definition given to machine precision, the calculation of 
the eigenvalues $\alpha_k$ is a possible
source of numerical inaccuracy. Note that
the error in these numbers is amplified in eq.~(8.8)
and may thus easily
affect the precision of the solution vector.

The matrix $M_{kl}$ should therefore be calculated using
a high-precision approximation to the Neuberger--Dirac operator.
It is also important to apply the operator $A$ in 
its original form, eq.~(8.1),
so that the eigenvalues $\alpha_k$ of the matrix are obtained with an error
equal to the approximation error on $\Dm$ times $(\alpha_k)^{1/2}$
(rather than just the approximation error).

\section 9. Adapted-precision inversion algorithm

Iterative improvement combined with 
the use of low-precision arithmetic
is mentioned in many books on numerical mathematics as 
an effective method to speed up the computation of 
the solution of linear systems
(see refs.~[\ref{GolubLoan},\ref{Recipes}] for example).
In the present context 
an additional acceleration can be
achieved by adjusting the accuracy of 
the approximation to the Neuberger--Dirac operator 
in the course of the inversion algorithm.
This has first been described in ref.~[\ref{GiustiMixedPrecision}],
and we now wish to present a second scheme that 
results in an even larger acceleration factor.

We again consider eq.~(8.1) in the chirality sector that does not
contain the zero-modes. The adapted-precision algorithm 
that will be discussed
should then be applied to the preconditioned system (8.7), but 
in order to keep the presentation as simple as possible
we shall stick to the original system in this section.

\subsection 9.1 Conjugate gradient algorithm

For the basic inversion algorithm we choose 
the standard conjugate gradient method.
To fix our notations we now first write down the
defining equations of this algorithm
even though these are very well known [\ref{GolubLoan},\ref{Recipes}].
 
The algorithm generates a sequence 
$\psi_i$, $i=1,2,3,\ldots$, of approximate
solutions recursively 
together with an accompanying sequence 
$p_i$ of so-called search directions. 
If we introduce the residues
\equation{
  g_i=\eta-A\psi_i,
  \enum
}
the recursion is given by
\equation{
  \psi_{i+1}=\psi_i+a_ip_i,\qquad\kern0.2cm
  a_i\equiv(g_i,g_i)/(p_i,Ap_i),
  \enum
  \next{2ex}
  g_{i+1}=g_i-a_iAp_i,
  \enum
  \next{2ex}
  p_{i+1}=g_{i+1}+b_ip_i,\qquad
  b_i\equiv(g_{i+1},g_{i+1})/(g_i,g_i),
  \enum
}
and is usually started by setting
\equation{
  \psi_1=0,\qquad g_1=\eta,\qquad p_1=g_1.
  \enum
}
Further information (on orthogonality and convergence properties
in particular) can be found
in the books quoted above.

\subsection 9.2 Iterative improvement

Once an approximate solution
$\chi_1$ of the linear system is found,
by performing a certain number of 
conjugate gradient iterations for example,
it can be improved in the following way.
First note that the norm of the residue
\equation{
  \eta_1=\eta-A\chi_1
  \enum
}
is much smaller than $\|\eta\|$ as otherwise we would
not consider $\chi_1$ to be an approximate solution.
Secondly it is obvious that the solution $\chi$ of
the subtracted system 
\equation{
  A\chi=\eta_1
  \enum
}
yields the solution of the original system through
\equation{
  \psi=\chi_1+\chi.
  \enum
}
The important point to note is that $\chi$ will in general 
be a small correction to $\chi_1$ since $\eta_1$ is small. 
In particular, if we solve
the subtracted system to a relative precision of 
a few decimal places, the accuracy of the total field $\psi$ 
is improved by these many digits.

We can now iterate this procedure and obtain the solution 
in the form
\equation{
  \psi=\sum_{k=1}^{\infty}\chi_k,
  \enum
  \next{2ex}
  \eta_0=\eta,\qquad\eta_k=\eta_{k-1}-A\chi_k\quad (k=1,2,3,\ldots),
  \enum
}
where the fields $\chi_k$ are assumed to be approximate solutions 
of the subtracted linear systems 
$A\chi=\eta_{k-1}$. Depending on how accurate the
solutions are, their magnitude $\|\chi_k\|$ will decrease
more or less rapidly,
but the algorithm is in any case always exact
since the inaccuracies are eventually corrected for.

\subsection 9.3 Computation of $\chi_k$

So far no acceleration has been achieved and we have merely reorganized
the calculation in a fairly complicated way.
The idea is now to use 
a numerical representation of $A$ with low precision 
when calculating $\chi_k$. In doing so we should of course be
very careful to choose a sufficiently accurate approximation
as otherwise it is possible that the conjugate gradient algorithm
becomes inefficient or even unstable. 

Before addressing this question, we recall that
$A$ acts in the subspace of negative chirality where
\equation{
  A=
  {2\over\abar}\,\Pm D_{m'}\Pm,
  \qquad
  m'=\frac{1}{2}\abar m^2.
  \enum
}
So if we use this representation of the operator and if 
a minmax polynomial is chosen that achieves a certain relative 
precision $\delta$, the 
total approximation error is
\equation{
  \bigl\|\widetilde{A}-A\bigr\|\leq
  \tau\equiv{2\over a^2}
  \bigl\{(1+s)^2-\frac{1}{4}(am)^2\bigr\}
  \bigl\{2(\kp+\km)+\delta\bigr\}.
  \enum
}
Note incidentally that 
the original form (8.1) of the operator offers no numerical advantage
at this point. 
The representation (9.11) 
is in fact more efficient by nearly a factor of $2$
for a given uniform approximation error
which is what counts here.

Now when the conjugate gradient algorithm is applied to solve the 
linear system $A\chi=\eta_{k-1}$,
a sequence $\chi_{k,i}$ of approximate solutions
is generated (cf.~subsect.~9.1).
The decrease of the associated residues
\equation{
  g_{k,i}=\eta_{k-1}-A\chi_{k,i}
  \enum
}
can then be observed and the recursion is stopped as soon as 
$\|g_{k,i}\|$ falls below a certain level or if the number of 
iterations reaches a specified maximal value. 
In the course of this calculation a minimal requirement
is that the residues are obtained to some precision at least, 
i.e.~the chosen approximation for the operator
$A$ should be such that the error $\tau\|\chi_{k,i}\|$ is 
always smaller than $\|g_{k,i}\|$. 

Since we cannot know in advance how large the approximate 
solutions $\chi_{k,i}$ are going to be, 
the accuracy $\tau$ should be adjusted dynamically
by increasing the degree of the minmax polynomial
when needed.
A relatively low precision, 
say $\delta=10^{-3}$, is certainly good enough for the first few iterations,
but as soon as $\tau\|\chi_{k,i}\|$
approaches $\|g_{k,i}\|$, the program should
switch to a better approximation. 
In general this may happen several times before
the recursion is halted and the calculated approximate solution $\chi_k$
is returned to the calling program.

\subsection 9.4 Computation of $\eta_k$ and global stopping criterion

The residues $\eta_k$ can in principle be calculated via eq.~(9.10),
but it is safer to obtain them directly from the 
current total solution vector through
\equation{
  \eta_k=\eta-A\,\sum_{l=1}^k\chi_l.
  \enum
}
In this way $\eta_k$ is guaranteed to be the true residue 
of the current approximation to the solution $\psi$ 
of the linear system that we wish to solve eventually.
In particular, if the global stopping criterion is based on the 
norm of the so computed residue, 
we need not worry about a possible
accumulation of approximation errors in the course of the 
calculation. The algorithm simply stops when and only when
an algorithm-independent criterion is fulfilled.

It is not difficult to figure out how good the 
numerical approximation of $A$ must be in eq.~(9.14).
Basically the approximation error $\tau$ times
the norm of the current total solution vector 
should be small compared to $\|\eta_k\|$ so that the residue
is correctly obtained to a few decimal places.
The stopping criterion for the
next subtracted system, $A\chi=\eta_k$, then has to 
be such that the accuracy of the solution is not driven
to a point, where the random digits in the source $\eta_k$
become important.

\subsection 9.5 Fine tuning of the algorithm

To achieve the maximal possible acceleration,
the degree of the minmax polynomial should 
never become very large in the course of the solution 
of the subtracted systems.
It is therefore important that the conjugate gradient algorithm
stops before this happens, which will be the case 
if the limits on the residue and the maximal number of iterations
are properly chosen.

A small additional improvement is obtained if the search direction
is passed from the current subtracted system
to the next as if the 
whole algorithm was a single sequence of conjugate gradient
iterations. 
This is a correct procedure up to rounding and approximation errors
that can partly be corrected by adjusting the initial 
search directions $p_{k,1}$ such that the orthogonality 
relations
\equation{
  \left((p_{k,1}-g_{k,1}),g_{k,1}\right)=
  \left((p_{k,1}-g_{k,1}),Ap_{k,1}\right)=0
  \enum
}
are restored to machine precision.

If the stopping criteria are well chosen,
this algorithm in general leads to significant savings 
in computer time
(factors of $3$ to $4$ are not uncommon).
As in the case of the calculation of the index 
of the Dirac operator discussed in sect.~6, the use of 
32~bit arithmetic can give another factor of $2$
or so on current PC processors [\ref{SSE}].
Single-precision arithmetic is in fact completely adequate 
for the approximate solution of the subtracted systems
[but will usually be insufficient
for the accumulation of the total solution vector
and the calculation of the residues (9.14)].

\section 10. Concluding remarks

Numerical simulations 
in the $\eps$--regime of QCD are difficult, not only because
they tend to require large amounts of computer time but 
also because special care needs to be taken to remain in control
of the approximation errors.
In this paper we have presented
an algorithmic framework that is fully satisfactory 
in this respect and that 
makes the calculations significantly faster.

Although we focused on a particular formulation of lattice QCD,
it is clear that many of the methods
that we have discussed are more widely applicable. 
An obvious case in this connection are the variants 
of lattice Dirac operators and gauge field actions
that have recently been studied in 
refs.~[\ref{DeGrandFatLink},\kern1pt%
\ref{ApproxFP}--\ref{OverlapImprovedAction}].
It is also conceivable that, under certain conditions,
even ordinary formulations of lattice QCD can profit
from mixed-precision algorithms and low-mode preconditioning.

A disadvantage of our algorithms is perhaps that 
the propagators for different quark masses cannot be obtained
simultaneously. In the $\eps$--regime the quark masses
are very small, however, and apart from the zero-mode contribution
(which must be calculated only once),
the quark propagator should not vary a lot from one value of the 
quark mass to another.
By taking linear combinations
of previously computed propagators as a first approximation,
the computer time needed for the calculation of the propagator
at the next quark mass should thus be significantly reduced.

The computation of the quark propagator in the $\eps$--regime,
using just the standard conjugate gradient algorithm,
requires a computational effort that grows
roughly like the square of the lattice volume at large volumes. 
Once low-mode preconditioning is switched on,
and if the number of subtracted
modes is scaled proportionally to some fractional power of the volume,
the situation becomes less transparent, but 
it is possible that a more favourable 
asymptotic behaviour will result.
In any case the total acceleration factor achieved by the 
methods described in this paper is quite impressive,
and we hope that physically interesting computations 
in the $\eps$--regime will now be feasible
on much larger lattices than previously considered.

\vskip1ex

We are indebted to Pilar Hern\'andez, Mikko Laine, Laurent Lellouch and
Peter Weisz for helpful discussions on chiral perturbation theory
and on the distribution of the low-lying eigenvalues of the Dirac
operator. We also wish to thank DESY for computer time 
and the staff of the computer centre for their support.
L.G.~was supported in part by the EU under
contract HPRN-CT-2000-00145.
C.H.~acknowledges support from the EU under 
contracts HPMF-CT-2001-01468 and HPRN-CT-2000-00145.

\appendix A. Notations

\vskip-2.5ex

\subsection A.1 Index conventions and Dirac matrices

Lorentz indices are taken from the middle of the Greek alphabet
and run from $0$ to $3$. Where appropriate repeated indices
are automatically summed over.
The Dirac matrices are assumed to satisfy
\equation{
  (\dirac{\mu})^{\dagger}=\dirac{\mu},\qquad
  \{\dirac{\mu},\dirac{\nu}\}=2\delta_{\mu\nu},
  \enum
}
but are otherwise left unspecified.
It is often advantageous, however, 
to choose a chiral representation where
\equation{
  \dirac{5}=\dirac{0}\dirac{1}\dirac{2}\dirac{3}
  =\pmatrix{1&0\cr 0&-1\cr}.
  \enum
}
In particular, pairs of Weyl fields (such as $e_k$ and $r_k$ in sect.~8)
can then conveniently 
be stored in the memory space allocated for a  Dirac field.

\subsection A.2 Wilson-Dirac operator

In terms of the gauge-covariant forward and backward difference
operators
\equation{
  \nab{\mu}\psi(x)={1\over a}
  \left\{U(x,\mu)\psi(x+a\hat{\mu})-\psi(x)\right\},
  \enum
  \next{2ex}
  \nabstar{\mu}\psi(x)={1\over a}
  \left\{\psi(x)-U(x-a\hat{\mu},\mu)^{-1}\psi(x-a\hat{\mu})\right\},
  \enum
}
the Wilson-Dirac operator is given by
\equation{
  D_{\rm w}=\frac{1}{2}\left\{
  \dirac{\mu}(\nabstar{\mu}+\nab{\mu})-a\nabstar{\mu}\nab{\mu}\right\},
  \enum
}
where $a$ denotes the lattice spacing,
$U(x,\mu)\in\SUthree$ the link variables and
$\hat{\mu}$ the unit vector in direction $\mu$.

\subsection A.3 Chebyshev polynomials

The Chebyshev polynomials $T_n(z)$, $n=0,1,2,\ldots$,
are defined through
\equation{
  T_n(\cos\theta)=\cos(n\theta).
  \enum
}
In particular, $T_0(z)=1$ and $T_1(z)=z$
while for larger degrees the polynomials
may be obtained algebraically using
\equation{
  T_{n+1}(z)-2zT_n(z)+T_{n-1}(z)=0,
  \quad n\geq1.
  \enum
}
The Clenshaw recursion [\ref{Recipes}] is a direct consequence of these
equations.

\appendix B. Proof of lemma~5.1

We first note that 
\equation{
  \|(1-\Pr)Q\Pr\|=\|\Pr Q(1-\Pr)\|=\varrho.
  \enum
}
This is a direct consequence of the definition (5.2) of $\varrho$
and of the hermiticity of the operators $Q$ and $\Pr$.
If we define the projected operator
\equation{
  \widetilde{Q}=\Pr Q\Pr+(1-\Pr)Q(1-\Pr),
  \enum
}
it follows that
\equation{
  \|Q-\widetilde{Q}\|=\|(1-\Pr)Q\Pr+\Pr Q(1-\Pr)\|=\varrho
  \enum
}
since $\Pr$ and $1-\Pr$ project to orthogonal subspaces.
The min-max principle (ref.~[\ref{ReedSimon}], Theorem XIII.1) 
now implies that the eigenvalues of $Q$ and 
$\widetilde{Q}$, ordered in ascending order and counting multiplicities, 
are by no more than a distance $\varrho$ apart.

\appendix C. Proof of lemma~5.2

We focus on the case of $\Prp$ since the argument is exactly
the same for the other projector.
So let us consider an eigenvalue $\nu_k>0$ and let
$u_k\in V$ be the associated normalized eigenvector of $\Pr Q\Pr$
[eq.~(5.3)].
From the definition (5.5) of $\varrho_k$ we then infer that
\equation{
  (Q-\nu_k)u_k=w_k,
  \qquad
  \|w_k\|=\varrho_k.
  \enum
}
Next we write down the equations 
\equation{
  \left\|\left[1-\Prpex\right]u_k\right\|^2=
  \left(u_k,\left[1-\Prpex\right]u_k\right)
  \noenum
  \next{2.5ex}
  {\phantom{\left\|\left[1-\Prpex\right]u_k\right\|^2}}
  =\bigl((Q-\lambda)u_k,{1-\Prpex\over(Q-\lambda)^2}\,(Q-\lambda)u_k\bigr)
  \enum
}
that are trivially true as long as $\lambda$ is not in the spectrum of $Q$.
We can actually safely set $\lambda=\nu_k$ 
because the distance $d_k$ of $\nu_k$ from the spectrum in the 
range of the projector $1-\Prpex$ does not vanish.
The norm of the operator ratio in eq.~(C.2) is then at most $1/d_k^2$ 
and this leads to the inequality
\equation{
  \left\|\left[1-\Prpex\right]u_k\right\|\leq\varrho_k/d_k
  \enum
}
when eq.~(C.1) is taken into account.

Together with the representation (5.4) of the projector $\Prp$
in terms of the eigenvectors $u_k$, this bound implies
\equation{
   \|\left[1-\Prpex\right]\Prp v\|\leq
   \sum_{\nu_k>0}(\varrho_k/d_k)\left|(u_k,v)\right|
   \enum
}
for any fermion field $v$, from which we deduce that
\equation{
  \|\Prp\left[1-\Prpex\right]\|=
  \|\left[1-\Prpex\right]\Prp\|\leq\kp.
  \enum
}
This is not quite what the lemma says, but 
the inequality is in fact more restrictive than it seems, because
$\Prpex$ and $\Prp$ are, by definition, projectors of the same rank.
In particular, the trace of the difference
$\Diff\equiv\Prp-\Prpex$ vanishes,
and a few lines of algebra show that 
\equation{
  \Diff^2+\Diff=2\Prp\left[1-\Prpex\right]\Prp
  \noenum\next{2ex}
  {\phantom{\Diff^2-\Diff= }}
  +\Prp\left[1-\Prpex\right](1-\Prp)+(1-\Prp)\left[1-\Prpex\right]\Prp.
  \enum
}
The operator thus satisfies
\equation{
  \Diff^{\dagger}=\Diff,
  \qquad
  \Tr\,\Diff=0,
  \qquad
  \left\|\Diff^2+\Diff\right\|\leq \kp\left(1+2\kp\right)
  \enum
}
and consequently has to be of order $\kp$.

\topinsert
\vbox{
\vskip0.0cm
\epsfxsize=7.0cm\hskip2.5cm\epsfbox{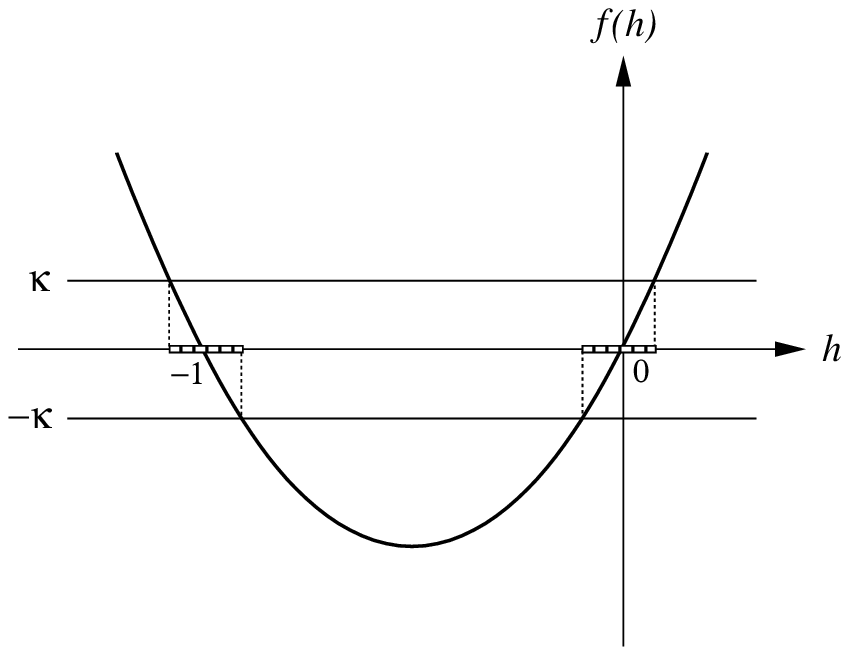}
\vskip0.4cm
\figurecaption{%
Plot of $f(h)=h^2+h$ and graphical determination of the 
ranges of $h$ (hashed bars) where the bounds (C.8) are satisfied.
}
\vskip0.0cm
}
\endinsert

To make this explicit, we set 
$\kappa=\kp\left(1+2\kp\right)$
and note that the eigenvalues $h$ of $\Diff$ have to be such that
\equation{
  -\kappa\leq h^2+h\leq\kappa
  \enum
}
(see fig.~2). In terms of $\kappa$ the condition stated on the 
last line of the lemma reads
\equation{
  2(l+1)\kappa<1
  \enum
}
so that $\kappa<\frac{1}{4}$ and
\equation{
  |h|\leq{\kappa\over1-2\kappa}\quad\hbox{or}\quad
  |h+1|\leq{\kappa\over1-2\kappa}.
  \enum
}
Now since $\Diff$ acts in a subspace of dimension $2l$ at most,
and since it has vanishing trace,
the second range in (C.10) is excluded if 
\equation{
  (2l-1){\kappa\over1-2\kappa}<1-{\kappa\over1-2\kappa}
  \enum
}
which is the case in view of (C.9).
All eigenvalues of $\Diff$ thus satisfy the first inequality in (C.10)
so that 
\equation{
  \|\Diff\|\leq{\kappa\over1-2\kappa}.
  \enum
}
Recalling the definition of $\Diff$ and $\kappa$, this proves the lemma.

\beginbibliography



\bibitem{GasserLeutwylerEpsI}
J. Gasser, H. Leutwyler,
Phys. Lett. B188 (1987) 477;
Nucl. Phys. B307 (1988) 763


\bibitem{NeubergerEps}
H. Neuberger,
Phys. Rev. Lett. 60 (1988) 889;
Nucl. Phys. B300 (1988) 180


\bibitem{HasenfratzLeutwylerEps}
P. Hasenfratz, H. Leutwyler,
Nucl. Phys. B343 (1990) 241


\bibitem{HansenEps}
F. C. Hansen,
Nucl. Phys. B345 (1990) 685

\bibitem{HansenLeutwyler}
F. C. Hansen, H. Leutwyler,
Nucl. Phys. B350 (1991) 201


\bibitem{LeutwylerSmilga}
H. Leutwyler, A. Smilga,
Phys. Rev. D46 (1992) 5607



\bibitem{GinspargWilson}
P. H. Ginsparg, K. G. Wilson,
Phys. Rev. D25 (1982) 2649


\bibitem{Kaplan}
D. B. Kaplan,
Phys. Lett. B288 (1992) 342;
Nucl. Phys. B (Proc. Suppl.) 30 (1993) 597

\bibitem{Shamir}
Y. Shamir,
Nucl. Phys. B406 (1993) 90

\bibitem{FurmanShamir}
V. Furman, Y. Shamir,
Nucl. Phys. B439 (1995) 54


\bibitem{Hasenfratz}
P. Hasenfratz,
Nucl. Phys. B (Proc. Suppl.) 63 (1998) 53;
Nucl. Phys. B525 (1998) 401

\bibitem{HLN}
P. Hasenfratz, V. Laliena, F. Niedermayer,
Phys. Lett. B427 (1998) 125


\bibitem{NeubergerDirac}
H. Neuberger,
Phys. Lett. B417 (1998) 141;
{\it ibid.}\/ B427 (1998) 353


\bibitem{ExactChSy}
M. L\"uscher,
Phys. Lett. B428 (1998) 342


\bibitem{Locality}
P. Hern\'andez, K. Jansen, M. L\"uscher,
Nucl. Phys. B552 (1999) 363


\bibitem{EdwardsHellerNarayanan}
R. G. Edwards, U. M. Heller, R. Narayanan,
Phys. Rev. D59 (1999) 094510

\bibitem{HernandezJansenLellouch}
P. Hern{\'a}ndez, K. Jansen, L. Lellouch,
Phys. Lett. B469 (1999) 198

\bibitem{DeGrandFatLink}
T. DeGrand (MILC collab.),
Phys. Rev. D63 (2001) 034503

\bibitem{HasenfratzHauswirthEtAl}
P. Hasenfratz, S. Hauswirth, T. J\"org, F. Niedermayer, K. Holland,
Nucl. Phys. B643 (2002) 280


\bibitem{ShuryakVerbaarschot}
E. V. Shuryak, J. J. Verbaarschot,
Nucl. Phys. A560 (1993) 306

\bibitem{VerbaarschotZahed}
J. J. Verbaarschot, I. Zahed,
Phys. Rev. Lett. 70 (1993) 3852

\bibitem{Verbaarschot}
J. J. Verbaarschot,
Phys. Rev. Lett. 72 (1994) 2531


\bibitem{DamgaardKernel}
P. H. Damgaard,
Phys. Lett. B424 (1998) 322


\bibitem{NishigakiDamgaardWettig}
S. M. Nishigaki, P. H. Damgaard, T. Wettig,
Phys. Rev. D58 (1998) 087704


\bibitem{OsbornToublanVerbaarschot}
J. C. Osborn, D. Toublan, J. J. Verbaarschot,
Nucl. Phys. B540 (1999) 317

\bibitem{DamgaardOsbornToublanVerbaarschot}
P. H. Damgaard, J. C. Osborn, D. Toublan, J. J. Verbaarschot,
Nucl. Phys. B547 (1999) 305


\bibitem{DamgaardNishigaki}
P. H. Damgaard, S. M. Nishigaki,
Phys. Rev. D63 (2001) 045012


\bibitem{GolubLoan}
G. H. Golub, C. F. van Loan,
Matrix computations, 2nd ed.
(Johns Hopkins University Press, Baltimore, 1989)



\bibitem{Bunk}
B. Bunk,
Nucl. Phys. B (Proc. Suppl.) 63 (1998) 952


\bibitem{NeubergerRational}
H. Neuberger,
Phys. Rev. Lett. 81 (1998) 4060;
Int. J. Mod. Phys. C 10 (1999) 1051


\bibitem{EdwardsRational}
R. G. Edwards, U. M. Heller, R. Narayanan,
Nucl. Phys. B540 (1999) 457


\bibitem{vandenEshofRational}
J. van den Eshof, A. Frommer, T. Lippert, K. Schilling, H. A. van der Vorst,
Comput. Phys. Commun. 146 (2002) 203


\bibitem{DamgaardHernandezEtAl}
P. H. Damgaard, P. Hern{\'a}ndez, K. Jansen, M. Laine, L. Lellouch,
Finite-Size Scaling of Vector and Axial Current Correlators,
hep-lat/0211020


\bibitem{Weinberg}
S. Weinberg,
Phys. Rev. D8 (1973) 605 and 4482

\bibitem{GaillardLee}
M. K. Gaillard, B. W. Lee,
Phys. Rev. Lett. 33 (1974) 108

\bibitem{AltarelliMaiani}
G. Altarelli, L. Maiani,
Phys. Lett. B52 (1974) 351

\bibitem{MaianiEtAl}
L. Maiani, G. Martinelli, G. C. Rossi, M. Testa,
Nucl. Phys. B289 (1987) 505

\bibitem{CapitaniGiusti}
S. Capitani, L. Giusti,
Phys. Rev. D64 (2001) 014506


\bibitem{BernardEtAl}
C. W. Bernard, T. Draper, A. Soni, H. D. Politzer, M. B. Wise,
Phys. Rev. D32 (1985) 2343


\bibitem{PilarMikko}
P. Hern{\'a}ndez, M. Laine, 
in preparation


\bibitem{DamgaardReview}
P. H. Damgaard,
Nucl. Phys. B (Proc. Suppl.) 106 (2002) 29


\bibitem{EdwardsHellerKiskisNarayanan}
R. G. Edwards, U. M. Heller, J. E. Kiskis, R. Narayanan,
Phys. Rev. Lett. 82 (1999) 4188;
Phys. Rev. D61 (2000) 074504

\bibitem{DamgaardEdwardsHellerNarayanan}
P. H. Damgaard, R. G. Edwards, U. M. Heller, R. Narayanan,
Phys. Rev. D61 (2000) 094503




\bibitem{Handscomb}
D. C. Handscomb (ed.),
Methods of numerical approximation
(Pergamon Press, Oxford, 1966)


\bibitem{Recipes}
W. H. Press, S. A. Teukolsky, W. T. Vetterling, B. P. Flannery,
Numerical recipes in FORTRAN, 2nd ed. (Cambridge University Press,
Cambridge, 1992)


\bibitem{EvaI}
B. Bunk, K. Jansen, M. L\"uscher, H. Simma,
Conjugate gradient algorithm to compute the low-lying
eigenvalues of the Dirac operator in lattice QCD,
notes (September 1994)

\bibitem{EvaII}
T. Kalkreuter, H. Simma,
Comput. Phys. Commun. 93 (1996) 33


\bibitem{SSE}
M. L\"uscher,
Nucl. Phys. B (Proc. Suppl.) 106 (2002) 21


\bibitem{ReedSimon}
M. Reed, B. Simon,
Methods of modern mathematical physics, vol. IV
(Academic Press, New York, 1978)


\bibitem{DiamantiniEtAl}
P. H. Damgaard, M. C. Diamantini, P. Hern\'andez, K. Jansen,
Nucl. Phys. B629 (2002) 445


\bibitem{GiustiMixedPrecision}
L. Giusti, C. Hoelbling, C. Rebbi,
Phys. Rev. D64 (2001) 114508
[E: {\it ibid.} D65 (2002) 079903];
Nucl. Phys. B (Proc. Suppl.) 106 (2002) 739


\bibitem{ApproxFP}
P. Hasenfratz, S. Hauswirth, K. Holland, T. J\"org, F. Niedermayer,
U. Wenger,
Nucl. Phys. B (Proc. Suppl.) 94 (2001) 627;
Int. J. Mod. Phys. C12 (2001) 691


\bibitem{GattringerI}
C. Gattringer,
Phys. Rev. D63 (2001) 114501


\bibitem{GattringerII}
C. Gattringer, I. Hip, C. B. Lang,
Nucl. Phys. B597 (2001) 451


\bibitem{HupercubicDirac}
W. Bietenholz,
Nucl. Phys. B644 (2002) 223


\bibitem{OverlapImprovedAction}
T. DeGrand, A. Hasenfratz, T. G. Kovacs,
hep-lat/0211006

\endbibliography

\bye